\documentstyle[12pt,epsf]{article}

\catcode`@=11
\newcount\@tempcntc
\def\@citex[#1]#2{\if@filesw\immediate\write\@auxout{\string\citation{#2}}\fi
  \@tempcnta\z@\@tempcntb\m@ne\def\@citea{}\@cite{\@for\@citeb:=#2\do
    {\@ifundefined
       {b@\@citeb}{\@citeo\@tempcntb\m@ne\@citea\def\@citea{,}{\bf ?}\@warning
       {Citation `\@citeb' on page \thepage \space undefined}}%
    {\setbox\z@\hbox{\global\@tempcntc0\csname b@\@citeb\endcsname\relax}%
     \ifnum\@tempcntc=\z@ \@citeo\@tempcntb\m@ne
       \@citea\def\@citea{,}\hbox{\csname b@\@citeb\endcsname}%
     \else
      \advance\@tempcntb\@ne
      \ifnum\@tempcntb=\@tempcntc
      \else\advance\@tempcntb\m@ne\@citeo
      \@tempcnta\@tempcntc\@tempcntb\@tempcntc\fi\fi}}\@citeo}{#1}}
\def\@citeo{\ifnum\@tempcnta>\@tempcntb\else\@citea\def\@citea{,}%
  \ifnum\@tempcnta=\@tempcntb\the\@tempcnta\else
   {\advance\@tempcnta\@ne\ifnum\@tempcnta=\@tempcntb \else \def\@citea{--}\fi
    \advance\@tempcnta\m@ne\the\@tempcnta\@citea\the\@tempcntb}\fi\fi}
\catcode`@=12

\newcommand{\alt}{\mathrel{\raisebox{-.6ex}{$\stackrel{\textstyle<}{\sim}$}}}
\newcommand{\agt}{\mathrel{\raisebox{-.6ex}{$\stackrel{\textstyle>}{\sim}$}}}

\renewenvironment{thebibliography}[1]
 {\begin{list}{\arabic{enumi}.}
    {\usecounter{enumi} \setlength{\parsep}{0pt}
     \setlength{\itemsep}{3pt} \settowidth{\labelwidth}{#1.}
     \sloppy
    }}{\end{list}}

\begin{document}
\font\fortssbx=cmssbx10 scaled \magstep2
\hbox to \hsize{
\hskip.5in \raise.1in\hbox{\fortssbx University of Wisconsin - Madison}
\hfill\vbox{\hbox{\bf MAD/PH/826}
            \hbox{March 1994}} }

\vglue1.5cm

\begin{center}
{\bf  IMPLICATIONS OF SUPERSYMMETRIC GRAND UNIFICATION}\footnote{%
Talk presented by V. Barger at the {\it International Conference on Unified
Symmetry --- in the Small and the Large}, Coral Gables, Florida, January
27--30, 1994.}\\[.5cm]
{\small V.~Barger, M.~S.~Berger, P.~Ohmann}\\[.2cm]
{\small\it Physics Department, University of Wisconsin, Madison, WI 53706, USA}
\end{center}

\vglue.6cm

\begin{abstract}\noindent
 Unification of the interactions of the Standard Model is possible in its
simplest supersymmetric extension.  The implications of the $\lambda _t$
fixed-point solution on the top mass and on Higgs phenomenology is discussed.
Expected correlations between the masses of various
supersymmetric particles are detailed.
\end{abstract}

\vglue .6cm
{\bf\noindent 1. Introduction}
\vglue 0.2cm
The search for symmetries beyond those in the Standard Model is a constant
task in  modern particle physics. Since
there is no compelling disagreement between the Standard Model and experiment,
why then should one look for the physics beyond the Standard Model? The
compelling reason is that the
mechanism behind electroweak symmetry breaking is completely unknown.

The attempts to describe the electroweak symmetry breaking of the Standard
Model fall largely into two broad classes: a weakly-interacting symmetry
breaking sector and a strongly-interacting
symmetry breaking sector.
What is a natural value for the mass of the Higgs bosons that characterize
the first case? It is easy to describe the requirements on the Higgs
sector in the minimal version of the supersymmetric standard model, commonly
referred to by its acronym MSSM. In this case the constraints from
supersymmetry on the Higgs sector leads to an upper bound on the mass of
the lightest physical Higgs boson.

The improvement in the precision data from LEP calls for a reevaluation
of the viability of grand unified theories in the context of supersymmetry.
Research has concentrated
recently on including two-loop contributions in the renormalization
group equations, the investigation of the impact of threshold corrections at
both the electroweak and grand unified scales, and the estimation of the
effects of non-renormalizable operators at the GUT scale.

\goodbreak
\vskip 0.6cm
{\bf\noindent 2. Phenomenological Motivations for
Supersymmetry\cite{susy}}
\vglue 0.2cm
The major motivations for supersymmetry are the following
\begin{itemize}
\item Unification of couplings\cite{susygut} ---
With SUSY, couplings evolve to an
intersection at $M \sim 10^{16}$ GeV. In the standard model, the gauge
coupling ``triangle''
fails to close and unification of gauge couplings
cannot be rescued even by large threshold corrections. See Figures 1a, b.

\item The problems with technicolor ---
The problems that flavor changing neutral
currents (FCNCs) and the generation of fermion masses
pose for technicolor theories are well known. The simplest technicolor theories
have problems when confronted with precision measurements of radiative
corrections in the electroweak theory.

\item Dark Matter --- A candidate for cold dark matter of the Universe arises
naturally in supersymmetry. The lightest supersymmetric particle (LSP) is
absolutely stable if a certain symmetry (known as R-parity) exists.

\item Radiative Breaking of the electroweak symmetry\cite{nac}--\cite{kkrw} ---
The Higgs mechanism can be understood in the context of supersymmetric GUTs
as a negative contribution to a Higgs mass-squared by a large logarithm of
the ratio of the GUT to electroweak scales.

\item Proton Decay\cite{an,hmy} --- In the
context of grand unified theories the heavy states
mediate transitions between quarks and leptons, thus violating lepton and
baryon number conservation. Since the rates for these transitions are governed
in part by the mass of the GUT scale states, the sensitive searches for
proton decay can impose severe restrictions on GUT models. In fact the minimal
SU(5) model predicts proton decay  at a rate already excluded by experiment.
The supersymmetric models have a higher unification scale and the dimension
six operators that plague the non-supersymmetric models are suppressed, but the
situation is complicated by the introduction of dimension five operators in
supersymmetry.

\end{itemize}

\vglue 0.6cm
{\bf\noindent 3. Evolution of Couplings: RGE}
\vglue 0.2cm
The unification of gauge groups is not a radical idea. In fact this
idea has already been partially
realized in nature as the Standard Model. The
only radical idea introduced by many grand unified theories is that there
is a ``desert'' from the electroweak scale upward to almost the Planck scale.

The gauge couplings evolve according to ordinary differential equations
derived from renormalization group ideas. Large logarithms which depend on
the scale of a process can be absorbed
into the gauge couplings. This gives rise to a ``running'' gauge coupling
that depends on scale. At the one-loop level these equations are not coupled
to each other, and the solutions for the reciprocal of the parameters
$\alpha _i\equiv g_i^2/4\pi $ are just linear functions of $t=\ln (Q/M_G)$
where $Q$ is the running mass
scale and $M_G$ is the GUT unification mass,
\begin{equation}
\alpha _i^{-1}(Q)=\alpha _i^{-1}(M_G^{})-{{b_i}\over {2\pi}}t \;.
\end{equation}
At the two-loop level the gauge couplings obey the
RGE\cite{EJ,susyrge2,susyrge1},
\begin{equation}
{{dg_i}\over {dt}}={g_i\over{16\pi^2 }}\left [b_ig_i^2+{1\over {16\pi^2 }}
\left (\sum _{j=1}^3b_{ij}g_i^2g_j^2-\sum _{j=t,b,\tau}a_{ij}g_i^2
\lambda _j^2\right )\right ] \label{dgidt} \;,
\end{equation}
The quantities $b_i$, $b_{ij}$, and $a_{ij}$ are determined by the particle
content in the effective theory.

Although unification can be restored in the non-supersymmetric case
by adding extra Higgs doublets that change the evolution of the electroweak
gauge couplings $\alpha _1$ and $\alpha _2$, this also lowers
the scale $M_G^{}$ at which unification occurs and thereby exacerbates the
violation of the proton decay bound rather than solving it as in the
supersymmetric case.
Another possibility is to put an intermediate scale between the
GUT scale and the electroweak scale. Then gauge coupling
unification occurs at the expense of adding new physics at this intermediate
scale, thereby adding to the complexity of unification and making the
resulting theory less predictive.

\begin{center}
\hbox{$\vcenter{\hsize=2.9in
\epsfxsize=2.9in\hspace*{0in}\epsffile{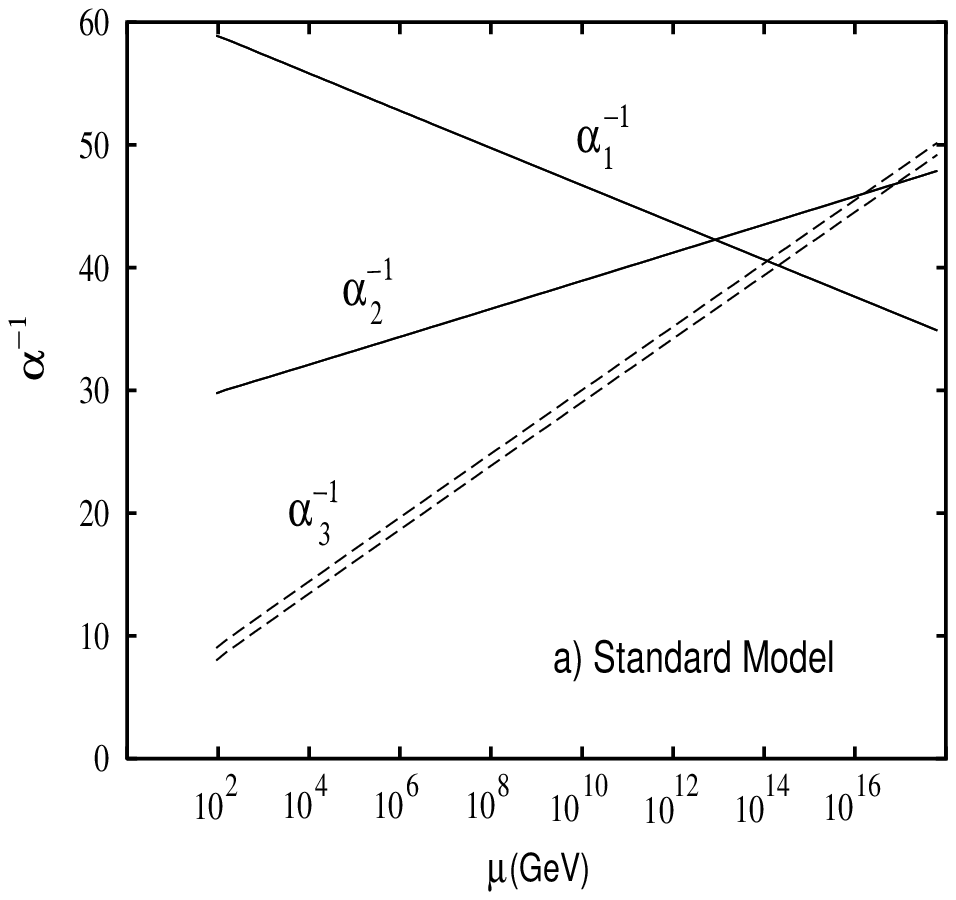}
}
\vcenter{\hsize=2.95in
\epsfxsize=2.95in\hspace*{.1in}\epsffile{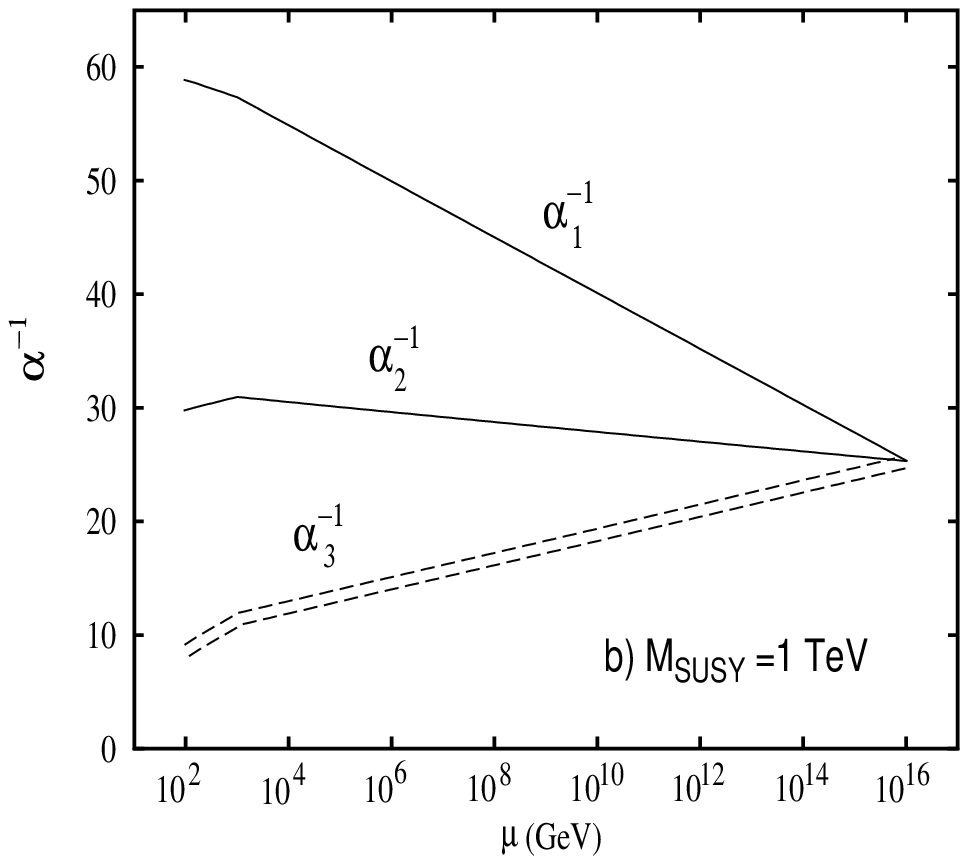}
}$}

\smallskip
{\small Fig.~1. Gauge coupling evolution (a) in  the SM and (b)~in
a SUSY-GUT example.}
\end{center}

\vglue 0.6cm
{\bf\noindent 4. Yukawa Coupling Evolution and the \boldmath{$\lambda_t$}
Fixed-Point Solution}
\vglue 0.2cm

The relationship between $m_b$ and $m_{\tau }$ in grand unified theories
incorporating a desert
is the most generic example of Yukawa coupling evolution.
It is interesting to note that this relationship can be made to work in
supersymmetric grand unified theories, but it implies a strong constraint on
the parameter space. In particular it gives a relationship between the top
quark mass and the angle $\beta $ that describes the alignment of the
vacuum in two Higgs doublet models (and the MSSM).

The Yukawa couplings are related to the fermions masses in our
convention\cite{bbo} by
\begin{eqnarray}\label{lambda_b}
\lambda_b(m_t) = {\sqrt2\, m_b(m_b)\over\eta_b v\cos\beta}\,, \qquad
\lambda_\tau(m_t) = {\sqrt2m_\tau(m_\tau)\over \eta_\tau v\cos\beta}\,, \qquad
\lambda_t(m_t) = {\sqrt2 m_t(m_t)\over v\sin\beta}\;.
\end{eqnarray}
The scaling factors $\eta _b$ and $\eta _{\tau }$ relate the Yukawa couplings
to their values at the scale $m_t$.
The evolution of these Yukawa couplings is described by the RGEs,
\begin{eqnarray}
{{d\lambda _t}\over {dt}}&=&{{\lambda _t}\over {16\pi^2}}\left [
-{13\over 15}g_1^2-3g_2^2-{16\over 3}g_3^2+6\lambda _t^2+\lambda _b^2\right ]
\;, \\
{{dR_{b/\tau}}\over {dt}}&=&{{R_{b/\tau}}\over {16\pi^2}}\left [
{4\over 3}g_1^2-{16\over 3}g_3^2+\lambda _t^2+3\lambda _b^2
-3\lambda _{\tau }^2\right ]\;.
\end{eqnarray}
where the ratio
$R_{b/\tau}\equiv {{\lambda _b}\over {\lambda _{\tau}}}$.
A well-known prediction of many GUT theories is that  $R_{b/\tau}$ is
equal to unity at the GUT scale\cite{ceg}
when the $b$ and $\tau$ are in
the same representation of the GUT gauge group. Figures 2 and 3 show
the solution
of these renormalization group equations for values of the bottom quark
mass. One sees that the top Yukawa coupling tends to be driven to its
infrared fixed point\cite{bbo},\cite{pendleton}--\cite{nmssm}.

\begin{center}
\epsfxsize=4.5in\hspace{0in}\epsffile{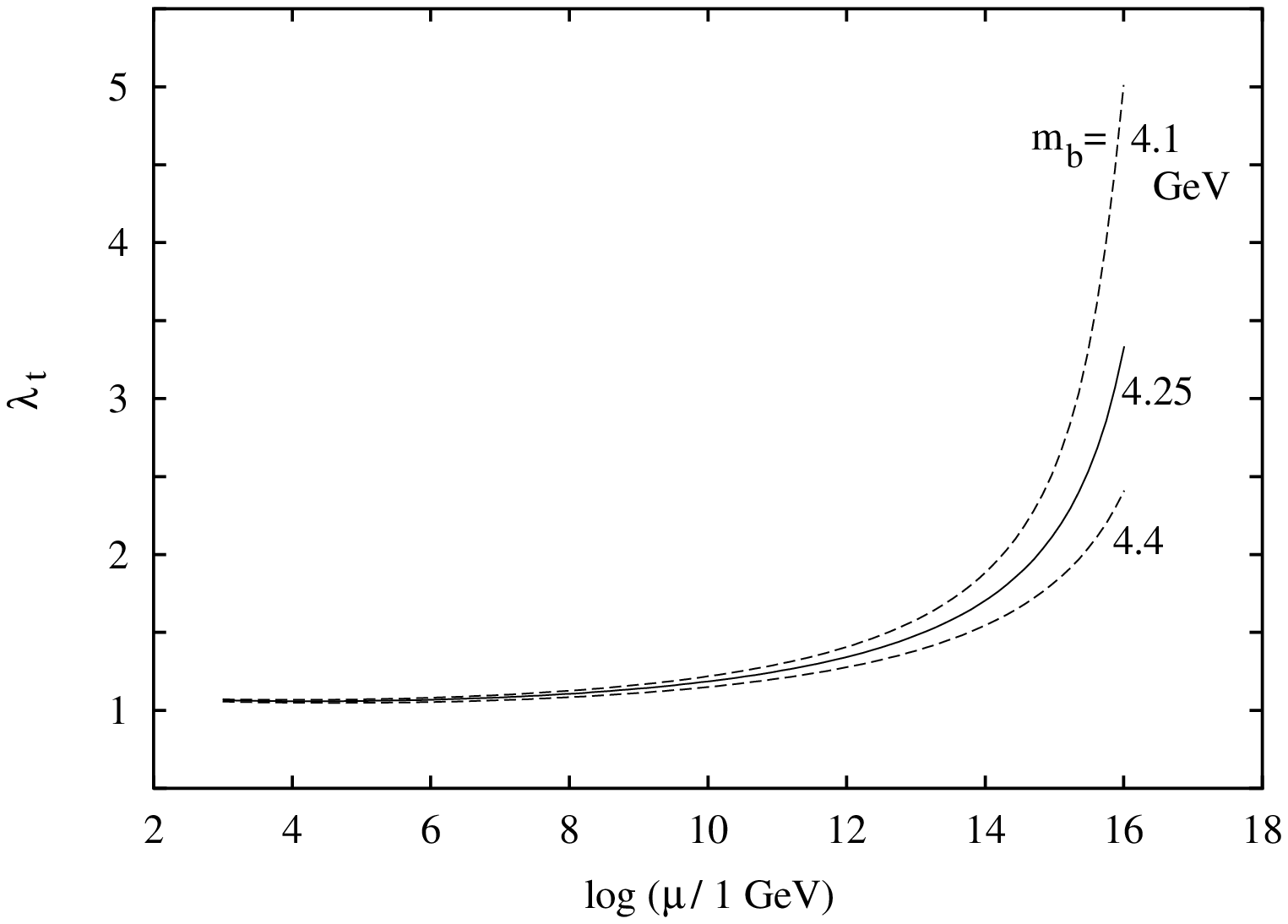}

\smallskip
\parbox{5.5in}{\small Fig.~2. If $\lambda _t$ is large at $M_G^{}$, then
the renormalization group equation causes $\lambda _t(Q)$ to evolve rapidly
towards an infrared fixed point as $Q \rightarrow m_t$ (from Ref.~\cite{bbo}).}
\end{center}

\smallskip

\begin{center}
\epsfxsize=4.75in\hspace*{0in}\epsffile{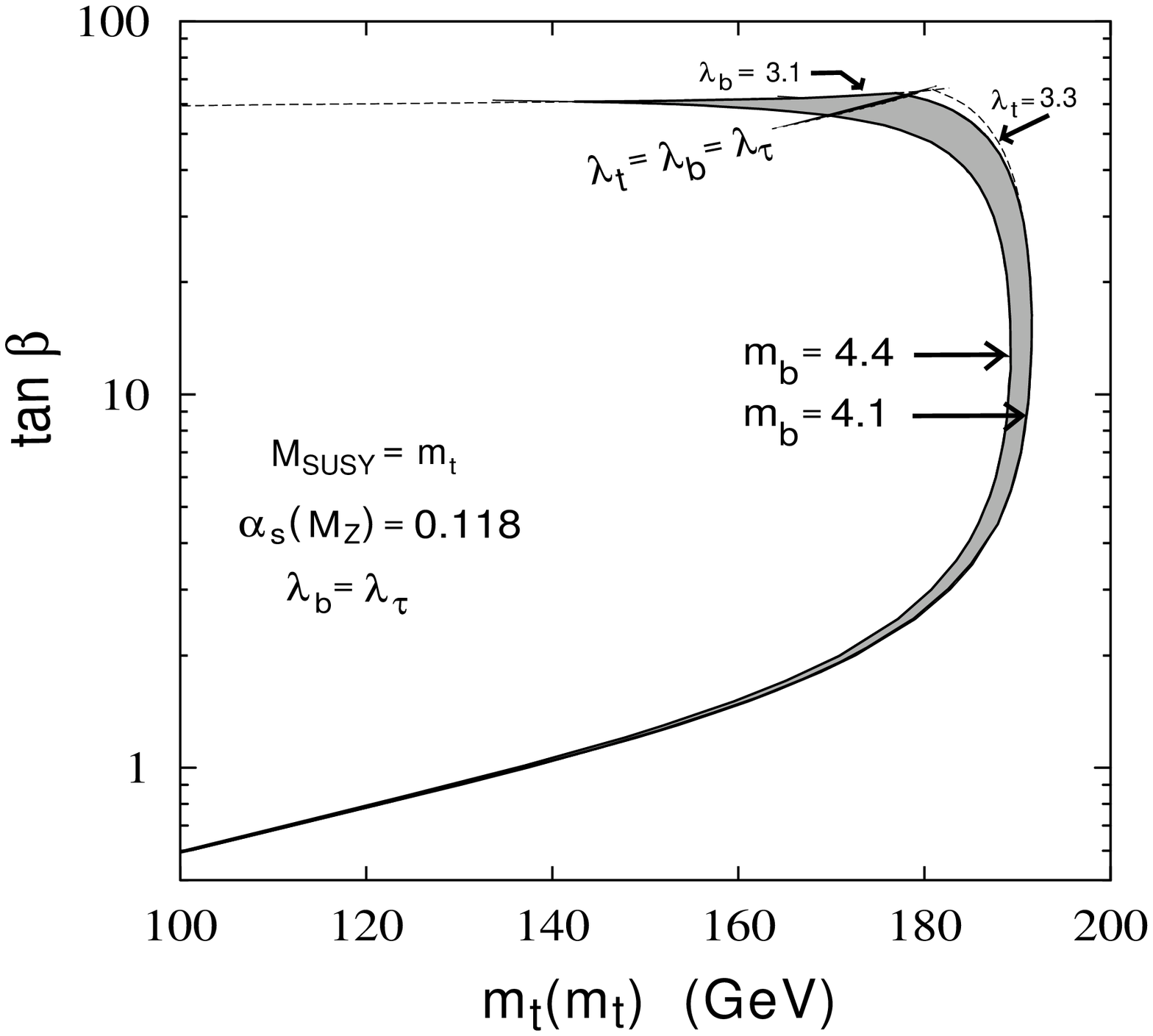}

\smallskip
\parbox{5.5in}{\small Fig.~3. Contours of constant $m_b(m_b)$ in the
$m_t(m_t),\tan\beta$ plane (from Ref.~\cite{bbo}).}
\end{center}

The fixed-point solution leads to the following
relation between the top quark mass (in the $\overline{DR}$ dimensional
reduction scheme\cite{dimred} with minimal
subtraction)  and $\tan \beta $.
\begin{eqnarray}
\lambda_t(m_t) &=& {\sqrt2 m_t(m_t)\over v\sin\beta} \qquad \Rightarrow
\qquad m_t(m_t)\approx {v\over {\sqrt{2}}}\sin \beta=(192 {\rm GeV})
\sin \beta \;
\end{eqnarray}
Converting this relation to the top quark pole mass yields\cite{bbo}
\begin{equation}
m_t^{\rm pole }\approx(200 {\rm GeV})\sin \beta \;.
\end{equation}

If one takes the $\lambda _t$
fixed-point solution and also assumes that the
top quark mass $m_t^{\rm pole}$ is less than about 160 GeV, important
consequences result for
the Higgs sector of the MSSM. From Fig.~4 it is clear that given these
assumptions $\tan \beta $ is very near one. Since $\tan \beta = 1$
is a flat direction
in the Higgs potential, for which the associated Higgs boson is massless at
tree level, and the true mass
of the lightest
Higgs is given almost entirely by the one-loop radiative corrections,
$m_h$ tends to be at the light end of its range.
This case was discussed in detail by Diaz and Haber\cite{dh}. In this low
$\tan \beta$ region
the Higgs mass is particularly sensitive to higher order
corrections\cite{kys}--\cite{lp3}. The upper bound
on $m_h$ that results is shown by the boundary of
the theoretically disallowed region in Fig.~5.

\begin{center}
\epsfxsize=4in\hspace*{0in}\epsffile{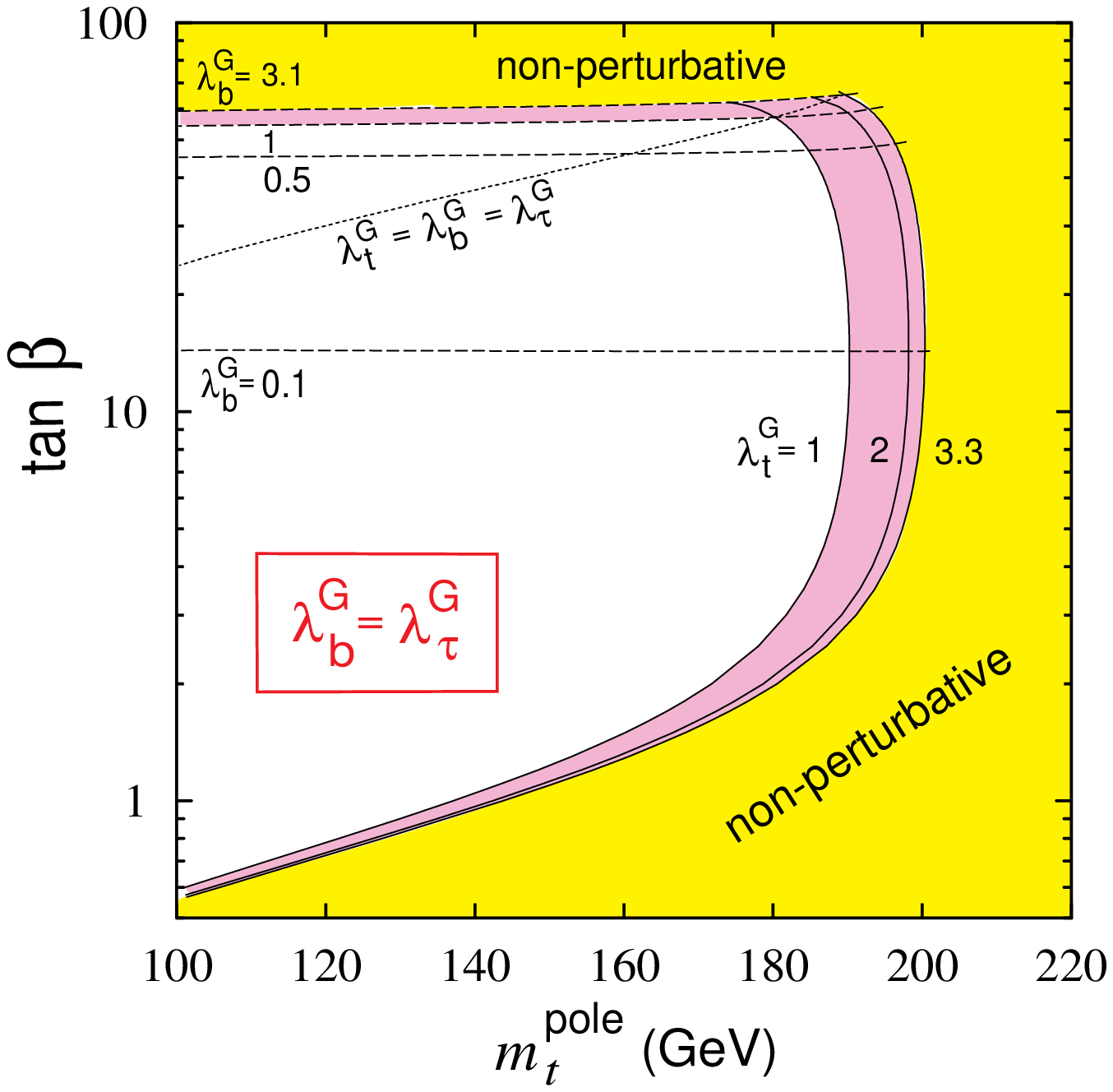}

\medskip
\parbox{5.5in}{\small Fig.~4. The fixed-point regions are given by Yukawa
couplings at the GUT scale being larger than about 1 ($\lambda _i^G\agt 1$).
Even larger values of the Yukawa couplings results in a breakdown of
perturbation theory.}
\end{center}

\begin{center}
\epsfxsize=4in\hspace{0in}\epsffile{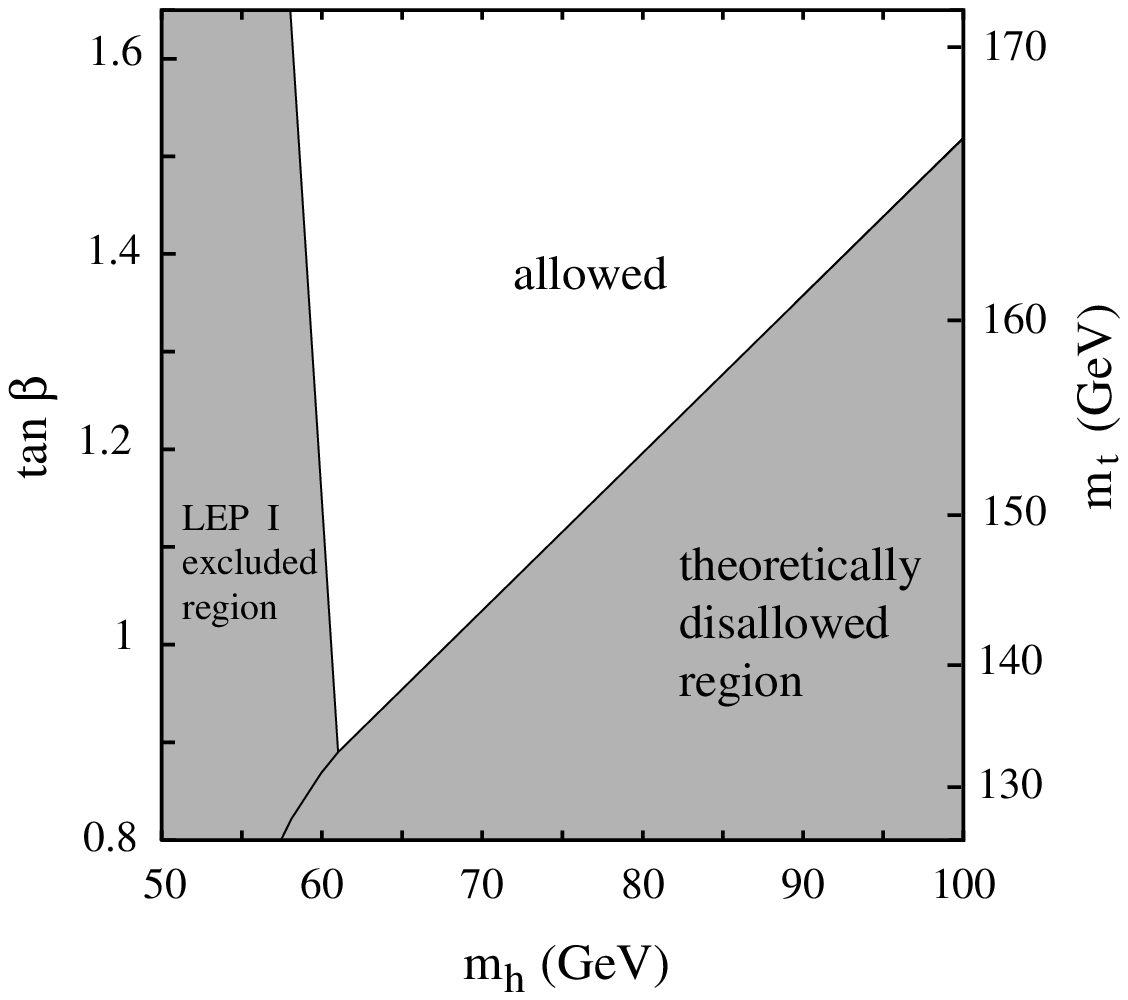}

\smallskip
\parbox{5.5in}{\small Fig.~5. The $\lambda_t$-fixed-point solution regions
allowed by the LEP\,I data in the $m_h, \tan \beta$ plane\cite{bbop}.
The top quark masses on the right hand side
are $m_t({\rm pole})$, correlated to
$\tan \beta $ by the fixed-point solution.}
\end{center}

If the top quark is sufficiently light, the fixed-point solution dictates that
the Higgs potential is such that the SUSY Higgs ($h$, $H$, $A$, $H^{\pm }$)
searches  are much more constrained.
The Higgs searches at LEP have excluded a Standard Model Higgs boson with
mass $m^{}_{H_{SM}^{}}$ less than 62.5 GeV.
The MSSM couplings give the relation
\begin{eqnarray}
\Gamma(Z\rightarrow Z^*h)=\sin ^2(\beta-\alpha )\Gamma(Z\rightarrow Z^*H_{SM})
\;.
\end{eqnarray}
for $m_h=m_{H_{SM}}^{}$.
Here the angle $\alpha $ describes the mixing between the CP-even components
of the two Higgs doublets. For a top quark mass less than about 170~GeV, the
fixed point gives $\tan \beta \approx 1$. Then in this region
of parameter space
\begin{eqnarray}
\sin ^2(\beta-\alpha )\sim 1\nonumber
\end{eqnarray}
so that the bound on the MSSM lightest Higgs is close to the bound on the
Standard Model Higgs.

\vglue 0.6cm
{\bf\noindent 5. Threshold Corrections}
\vglue 0.2cm
As discussed above, the requirement of gauge and Yukawa coupling unification
can be successful in the minimal versions of supersymmetric grand unified
theories and can result in strong constraints on the parameter space of
the model. With the improved electroweak data recently made available, a
number of groups have been attempting to incrementally refine the theoretical
predictions. The one-loop renormalization group equations have been extended
to two-loops in the Yukawa sector\cite{susyrge2,bbo}
and in the soft-supersymmetry breaking parameters\cite{mvy}.
Moreover threshold corrections at the GUT and electroweak
(and electroweak-scale SUSY thresholds) have been
investigated\cite{cpw,lp,lp3,bh,hs,hrs,hempfling,wright}.
The RGE evolution yields the generic features that result from the large
separation of scales typical in GUT theories.
Threshold effects are typically model-dependent and sensitive to the
detailed spectrum (the supersymmetric spectrum, the top mass, etc.\ at the
electroweak scale and the superheavy spectrum at the GUT scale)\footnote{The
Yukawa coupling unification condition $\lambda _b^G=\lambda _{\tau}^G$ is
itself a model-dependent feature satisfied in simple GUT scenarios.}.
Threshold effects have been incorporated into the predictions from gauge and
Yukawa coupling unification and in the Higgs potential,
and eventually will be included in the full supersymmetric
spectrum as well\cite{hmg}.

The theory below the grand unified theory is an effective theory with the
heavy GUT-scale particles integrated out. Since the heavy particles do not
completely fill the representations of the grand unified gauge group, the
group is broken and the RGEs of the effective symmetry exhibit this
broken symmetry and the gauge couplings diverge below the GUT scale.
The process of integrating out the heavy particles gives rise to threshold
corrections that depend on the details of the grand unified theory.
Since the threshold corrections at the
GUT scale depend on the superheavy spectrum,
one therefore expects these corrections to be constrained by the proton
decay limits.
In a similar way there are threshold corrections at the electroweak scale
from the effective theories that are introduced there (the scale of
supersymmetry can be chosen to be different than the electroweak scale).
Typically one makes some assumptions about the supersymmetric spectrum to
simplify the problem (such as in a supergravity-based scenario).
Ultimately one hopes that these model-dependent features can be used to
distinguish between the various realization of the supersymmetric models.

The threshold corrections to Yukawa coupling unification are also
relevant\cite{wright} to
the analyses of GUT scale mass matrix ans\"{a}tze. Relations
between fermion masses and mixing angles that arise in these scenarios
will be modified by model-dependent effects.

\vglue 0.6cm
{\bf\noindent 6. Where are the Sparticles?}
\vglue 0.2cm
Sometimes supersymmetry is criticized because of a
proliferation of parameters. This is not necessarily a fair criterion,
since in
some minimal versions of supergravity theories the complete mass spectrum
and couplings can be explained with the addition of as few as three or five
parameters. At the present time our ignorance of the mechanism
of supersymmetry breaking should make us cautious about sweeping statements
about the supersymmetric particles that rely on some GUT-scale assumptions;
however, it is not unreasonable to expect there
will be correlations between the supersymmetric particle masses and couplings
since we hope that the ultimate theory that describes them near the
Planck scale is a simple and economical one. Figure~6 shows representative
results for RGE evolution of the sparticle masses.

\begin{center}
\epsfxsize=5in\hspace*{0in}\epsffile{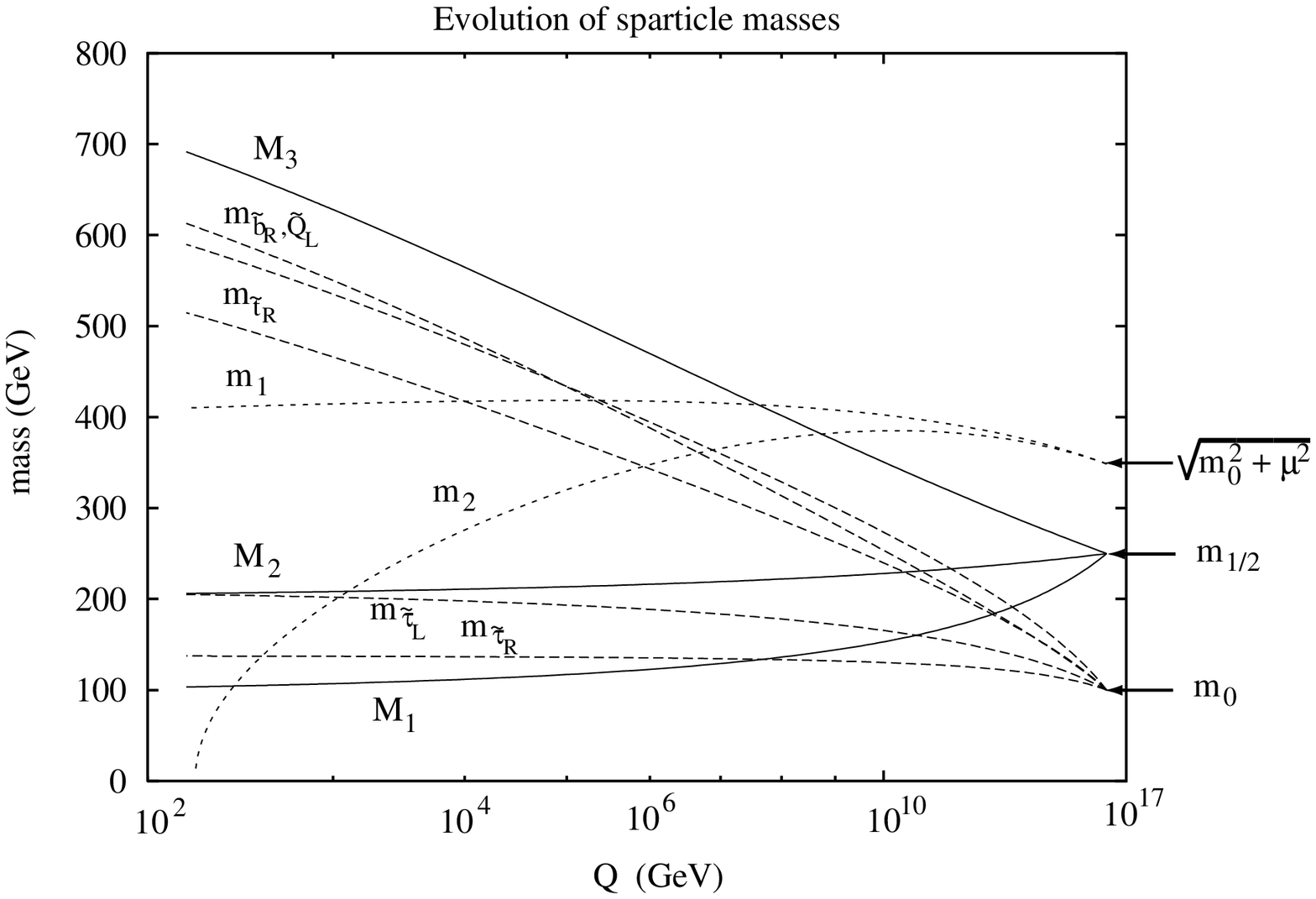}

\smallskip
\parbox{5.5in}{\small Fig.~6. The evolution of the sparticle spectrum from the
unification scale down to the electroweak scale. The characteristic
behavior exhibited by the mass parameters are typical
of renormalization group equation evolution.}
\end{center}

A popular and convenient approach\cite{Kelley} to obtaining a solution to the
soft-supersymmetry breaking RGEs is to define some inputs at the GUT scale and
some inputs at the electroweak scale. We have dubbed this approach the
ambidextrous approach\cite{madph801} to distinguish it from the
bottom-up\cite{op2} and top-down\cite{hempf2} approaches where all inputs are
defined at the same scale.
Common to all of these approaches is the requirement that correct electroweak
symmetry breaking (EWSB) be achieved. This is accomplished by imposing two
minimization conditions obtained by minimizing the Higgs potential.

The tree-level Higgs potential is given by
\begin{eqnarray}
V_0&=&(m_{H_1}^2+\mu ^2)|H_1|^2+(m_{H_2}^2+\mu
^2)|H_2|^2+m_3^2(\epsilon_{ij}{H_1}^i{H_2}^j+{\rm h.c.})
\nonumber \\
&&+{1\over 8}(g^2+g^{\prime 2})\left [|H_1|^2-|H_2|^2\right ]^2
+{1\over 2}g^2|H_1^{i*}H_2^i|^2\;, \label{tree}
\end{eqnarray}
where $m_{H_1}^{}$, $m_{H_2}^{}$, and
$m_3$ are soft-supersymmetry breaking parameters and $\epsilon _{ij}$ is the
total antisymmetric tensor.
The minimum of the Higgs potential must occur by the acquisition
of vacuum expectation values.
Minimizing $V_0$ with respect to the two neutral CP-even Higgs degrees of
freedom yields
\begin{eqnarray}
{1\over 2}M_Z^2&=&{{m_{H_1}^2-m_{H_2}^2\tan ^2\beta }
\over {\tan ^2\beta -1}}-\mu ^2 \;.  \label{treemin1} \\
-B\mu &=&{1\over 2}(m_{H_1}^2+m_{H_2}^2+2\mu ^2)\sin 2\beta \;.
\label{treemin2}
\end{eqnarray}
The masses in these equations are running masses that depend on the scale
$Q$ in the RGEs that describe their evolution. Hence the solutions
obtained are functions of the scale $Q$.
Equations~(\ref{treemin1}) and (\ref{treemin2}) are a particularly convenient
form since the gauge couplings dependence (the D-terms
in the language of supersymmetry) is isolated in Eq.~(\ref{treemin1}). This
equation also clearly shows the fine-tuning problem that may be present in
the radiative breaking of the electroweak symmetry. For large values of
$|\mu |$, there must be a cancellation between large terms on the right hand
side to obtain the correct experimentally measured $M_Z^{}$ (or equivalently
the electroweak scale). For $\tan \beta $ near one, a cancellation of
large terms must occur. Finally, these minimization equations illustrate the
power of the ambidextrous approach. For EWSB to be satisfied one need only
specify $m_t$, $M_Z$, $\tan \beta $ at the electroweak scale and the common
gaugino mass $m_{1/2}$, scalar mass $m_0$, and the trilinear scalar coupling
$A$ at the GUT scale. Then one solves the minimization equations given
above to obtain $\mu $ (up to a sign) and $B$, thereby implicitly
satisfying the EWSB requirement. For more details, see Ref.~\cite{madph801}.

A heavy top quark produces large corrections to the Higgs potential of the
MSSM\cite{msb}.
Gamberini, Ridolfi, and Zwirner showed\cite{grz} that the tree-level
Higgs potential is inadequate for the purpose of analyzing radiative breaking
of the electroweak symmetry because the tree-level Higgs vacuum expectation
values
$v_1$ and $v_2$ are very sensitive to the scale at which the renormalization
group equations are evaluated. The one-loop corrections to the
Higgs potential effectively moderates this sensitivity to the scale $Q$.
The one-loop corrections are conveniently calculated using the tadpole
method\cite{madph801,tadpole,sher}. The corrections to
Eqs.~(\ref{treemin1}) and
(\ref{treemin2}) can be obtained by calculating the two tadpoles
with two independent CP-even Higgs as external lines as in Fig.~7.
The one-loop corrected minimization conditions can then be used to generate
a complete supersymmetric particle spectrum which satisfies EWSB.

\begin{center}
\epsfxsize=.75in\hspace*{0in}\epsffile{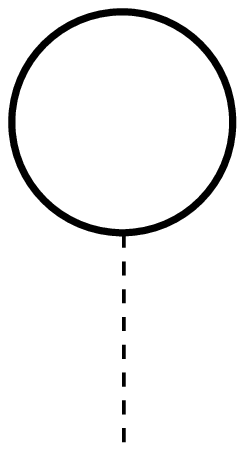}

\smallskip
\parbox{5.5in}{\small Fig.~7. The tadpole diagrams offer a simple method
to obtain the one-loop modifications to the minimization conditions.
The loop consists of all matter and gauge-Higgs contributions, and the
external lines are the two CP-even Higgs fields.}
\end{center}

\medskip
Including only the leading contribution coming from the top quark loop (and
neglecting the D-term contributions to the squark masses)
one obtains the expressions
\begin{eqnarray}
{1\over 2}M_Z^2&=&{{m_{H_1}^2-m_{H_2}^2\tan ^2\beta }
\over {\tan ^2\beta -1}}-\mu ^2
-{{3g^2m_t^2}\over {32\pi ^2M_W^2\cos 2\beta}}\Biggl[ 2f(m_t^2)
-f(m_{\tilde{t}_{1}}^2)-f(m_{\tilde{t}_{2}}^2) \nonumber\\
&& \hspace{2.5in} {} +{{f(m_{\tilde{t}_{1}}^2)
-f(m_{\tilde{t}_{2}}^2)}
\over {m_{\tilde{t}_1}^2-m_{\tilde{t}_2}^2}}
\Bigl( (\mu \cot \beta)^2-A_t^2\Bigr) \Biggr] \;, \nonumber\\
\label{treemin3}\\
-B\mu &=&{1\over 2}(m_{H_1}^2+m_{H_2}^2+2\mu ^2)\sin 2\beta
-{{3g^2m_t^2\cot \beta }\over {32\pi ^2M_W^2}}\Biggl[ 2f(m_t^2)
-f(m_{\tilde{t}_{1}}^2)-f(m_{\tilde{t}_{2}}^2) \nonumber\\
&& \hspace{1.75in} {} -{{f(m_{\tilde{t}_{1}}^2)-f(m_{\tilde{t}_{2}}^2)}
\over {m_{\tilde{t}_1}^2-m_{\tilde{t}_2}^2}}
(A_t+\mu \cot \beta )(A_t+\mu \tan \beta )\Biggr]\;, \nonumber \\
\label{treemin4}
\end{eqnarray}
where
\begin{eqnarray}
f(m^2)=m^2\left (\ln {{m^2}\over {Q^2}}-1\right )\;.
\end{eqnarray}
The extra one-loop contribution included above renders the solution less
sensitive to the scale $Q$\cite{aspects,op2,Ramond,madph801,cc},
as can be shown explicitly by examining the
relevant renormalization group equations for the parameters that enter into
the minimization conditions.
The complete expressions for the one-loop contributions can be found in
Ref.~\cite{madph801}.
The fine-tuning problem is alleviated somewhat, but not entirely, by the
inclusion of one-loop corrections to the Higgs potential.
As our naturalness criterion we require
\begin{eqnarray}
&&|\mu (m_t)|< 500\ {\rm GeV}\;.
\end{eqnarray}

The gaugino masses are related (through one-loop
order) by the same ratios that describe the gauge couplings at the
electroweak scale. This observation, together with the fact that $|\mu|$ is
large, yields simple correlations between the lightest chargino and neutralinos
and the gluino\cite{arnowittnath,Ramond,lnp}, namely
\begin{eqnarray}
M_{\chi _1^{0}}&\simeq &M_1 \;, \\
M_{\chi _1^{\pm}}&\simeq &M_{\chi _2^0}\;\;\simeq \;\;M_2\;\;=\;\;
{{\alpha _2}\over {\alpha_1}}M_1\;\simeq \;\;2M_1\;\;\simeq \;\;2M_{\chi _1^0}
\;, \\
m_{\tilde{g}}&=&M_3\;\;=\;\;{{\alpha _3}\over {\alpha_2}}M_2\;\;=\;\;
{{\alpha _3}\over {\alpha_1}}M_1\;.
\end{eqnarray}
In our analysis the quantities in these equations are evaluated at
scale $m_t$.
The heaviest chargino and the two heaviest neutralino states are primarily
Higgsino with
\begin{eqnarray}
M_{\chi _2^{\pm}}&\simeq &M_{\chi _3^0}\;\;\simeq \;\;M_{\chi _4^0}\;\;\simeq
\;\;|\mu | \;.
\end{eqnarray}

As previously noted the mass of the lightest Higgs $h$ arises mainly from
radiative corrections\cite{bbop,dh,lp3,lnpwz}. The heavy Higgs states are
(approximately) degenerate $\approx M_A$ because at tree-level
$M_A=-{{B\mu }\over {\sin 2\beta}}\approx -B\mu$ is large.
The squark and slepton masses also display
simple asymptotic behavior at large $|\mu|$.
The first and second squark generations are approximately degenerate.
The splitting of the stop quark masses grows as $|\mu|$ increases.
The splitting of the sbottom states does not change much with $\mu $ for
small $\tan\beta$.

The approximate experimental bounds that we impose
are listed in Table~1. Together with our naturalness criteria
$|\mu (m_t)|<500$~GeV, these bounds give the allowed region in the
$m_0,m_{1/2}$ plane  shown as the shaded areas in Fig.~8.

\bigskip

\begin{center}
{\small Table 1.  Approximate experimental bounds.}
\medskip
\begin{tabular}{|c|c|}
\hline
Particle& Experimental Limit (GeV)
\\ \hline \hline
gluino& 120
\\ \hline
squark, slepton& 45
\\ \hline
chargino& 45
\\ \hline
neutralino& 20
\\ \hline
light higgs& 60
\\ \hline
\end{tabular}
\end{center}

\begin{center}
\epsfxsize=4in\hspace*{0in}\epsffile{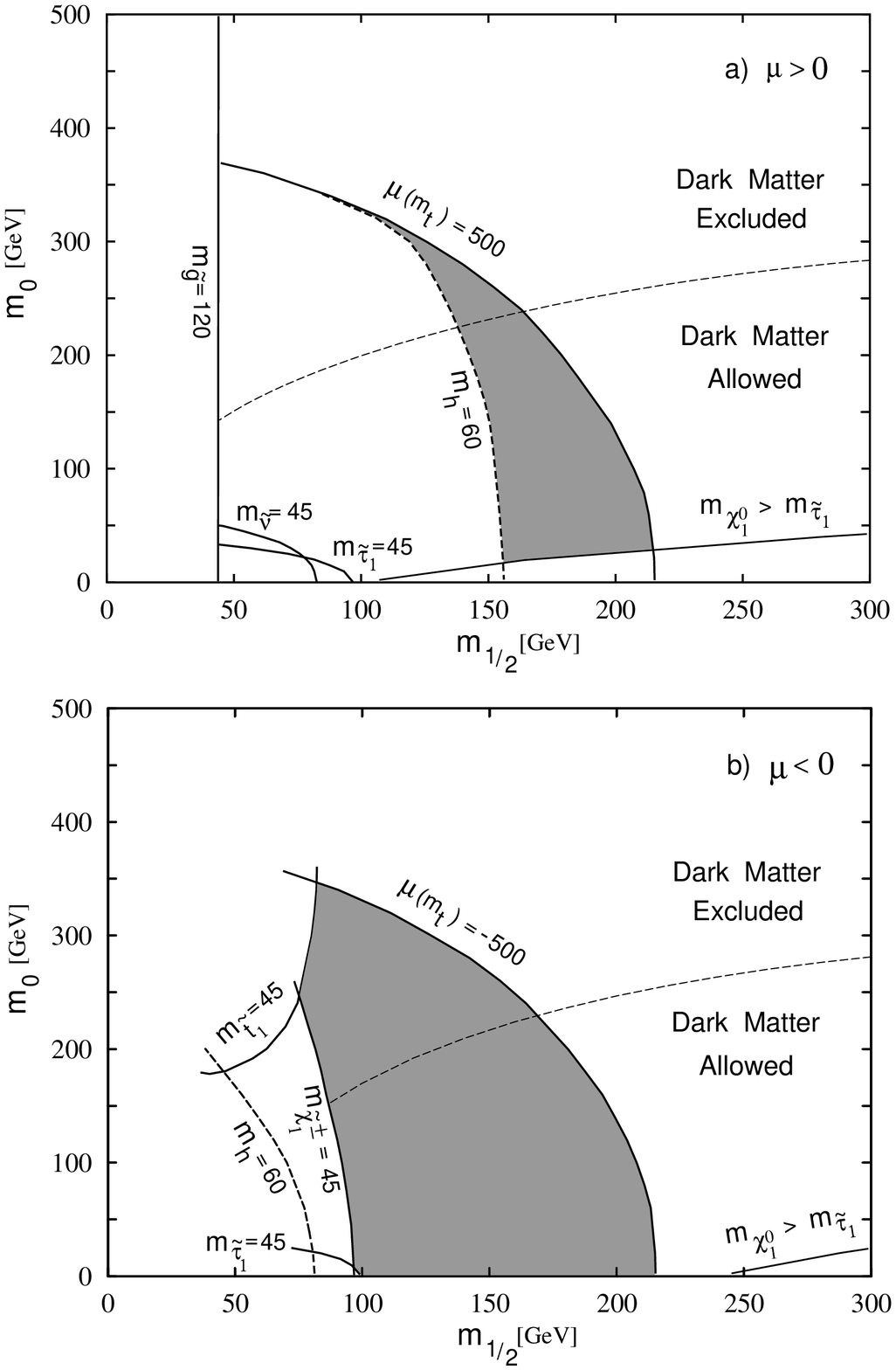}

\medskip
\parbox{5.5in}{\small Fig.~8. Allowed regions of parameter space for $m_t(m_t)
= 160$ GeV, $\tan \beta=1.47$ (a low-$\tan \beta$ fixed-point solution)
(from Ref.~\cite{madph801}).}
\end{center}

The prediction for $m_h$ in the low-$\tan \beta $ region is particularly
sensitive to two-loop corrections\cite{lp3}. Hence the precise location of the
$m_h=60$ GeV contour is somewhat uncertain. The MSSM has conserved R-parity so
the lightest supersymmetric particle (LSP) is stable. Usually the LSP is the
lightest neutralino, but for small values of $m_0$ the supersymmetric partner
of the tau lepton is sometimes lighter.
For the lightest SUSY particle to be neutral
there is an upper bound on the value of $m_{1/2}$ for small
$m_0$. In particular such an upper bound exists for no-scale models ($m_0=0$),
and is more stringent for $\mu >0$ due to the mixing between the left and right
handed $\tilde{\tau }$, giving a stau lighter than the lightest neutralino. The
LSP can also account for the dark matter of the
Universe\cite{dm,RoRo}.
The large values of $\mu $ obtained from the low $\tan\beta$ solution result in
the lightest neutralino being
predominantly gaugino (see Figure 9). This leads to a reduced rate of
annihilation of
neutralinos and can provide too much relic abundance and overclose the
Universe. This constraint is shown as the dashed line in Figure 8; this line
should be regarded as a semi-quantitative one only since the contributions of
$s$-channel poles that can enhance the annihilation rate have been neglected.

\begin{center}
\epsfxsize=3.35in\hspace*{0in}\epsffile{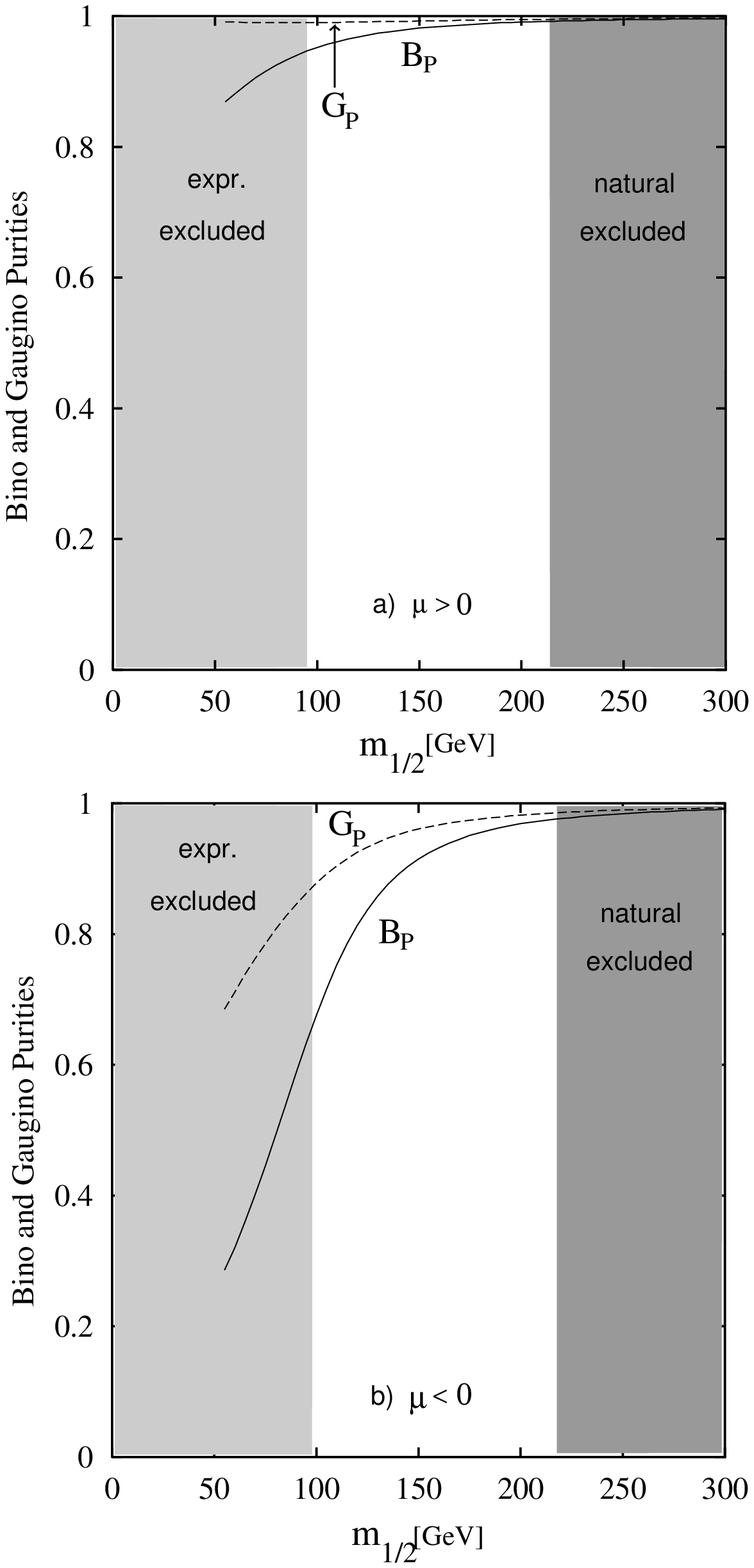}

\smallskip
{\small Fig.~9. Bino and gaugino purities for low $\tan \beta $
fixed-point solution.}
\end{center}

Figure 10 shows the dependence of the sparticle masses on the parameter
$m_{1/2}$. The large $|\mu |$ obtained from the low-$\tan \beta $ solutions
lead to highly correlated masses. Figure~11 shows the squark masses, which
are quite degenerate for the light families.

Figure 12 shows the supersymmetric particle mass dependence on the parameter
$m_0$ for  a fixed value of $m_{1/2}$. The lighter top squark eigenstate
$\tilde{t}_1$ has an approximately constant mass with increasing $m_0^{}$.
This occurs because
$|\mu | $ increases with $m_0^{}$ giving rise to increased mixing between
the left- and right-handed top squarks (lowering the lightest mass
eigenstate).

\begin{center}
\epsfxsize=4.5in\hspace*{0in}\epsffile{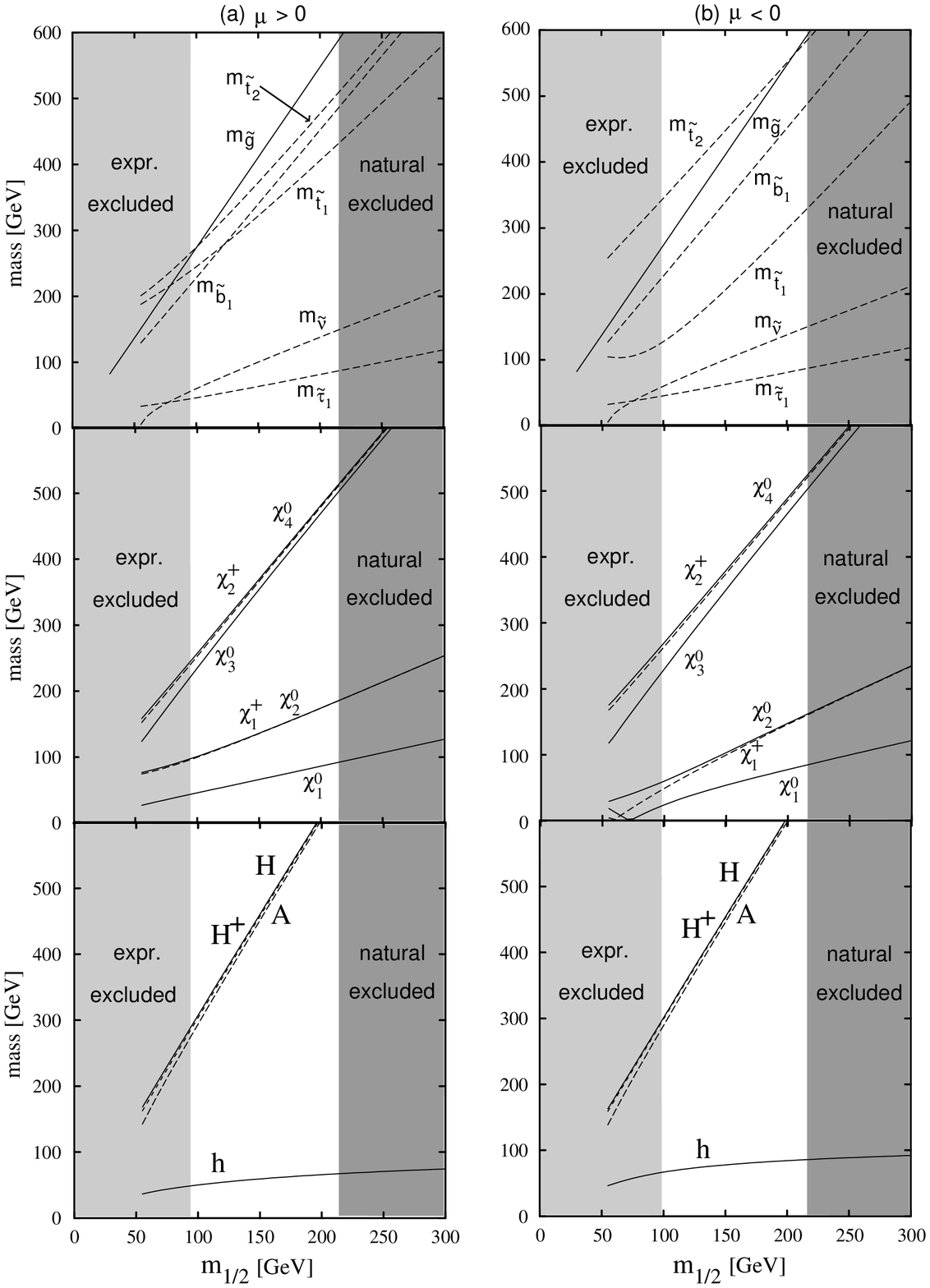}

\medskip
\parbox{5.5in}{\small Fig.~10. Particle spectra as a function of $m_{1/2}$. The
shaded bands at the left are regions excluded by the experimental constraints.
The shaded bands at the right labeled ``natural excluded'' require
$|\mu (m_t)| > 500$ GeV (from Ref.~\cite{madph801}).}
\end{center}

\begin{center}
\epsfxsize=3.5in\hspace*{0in}\epsffile{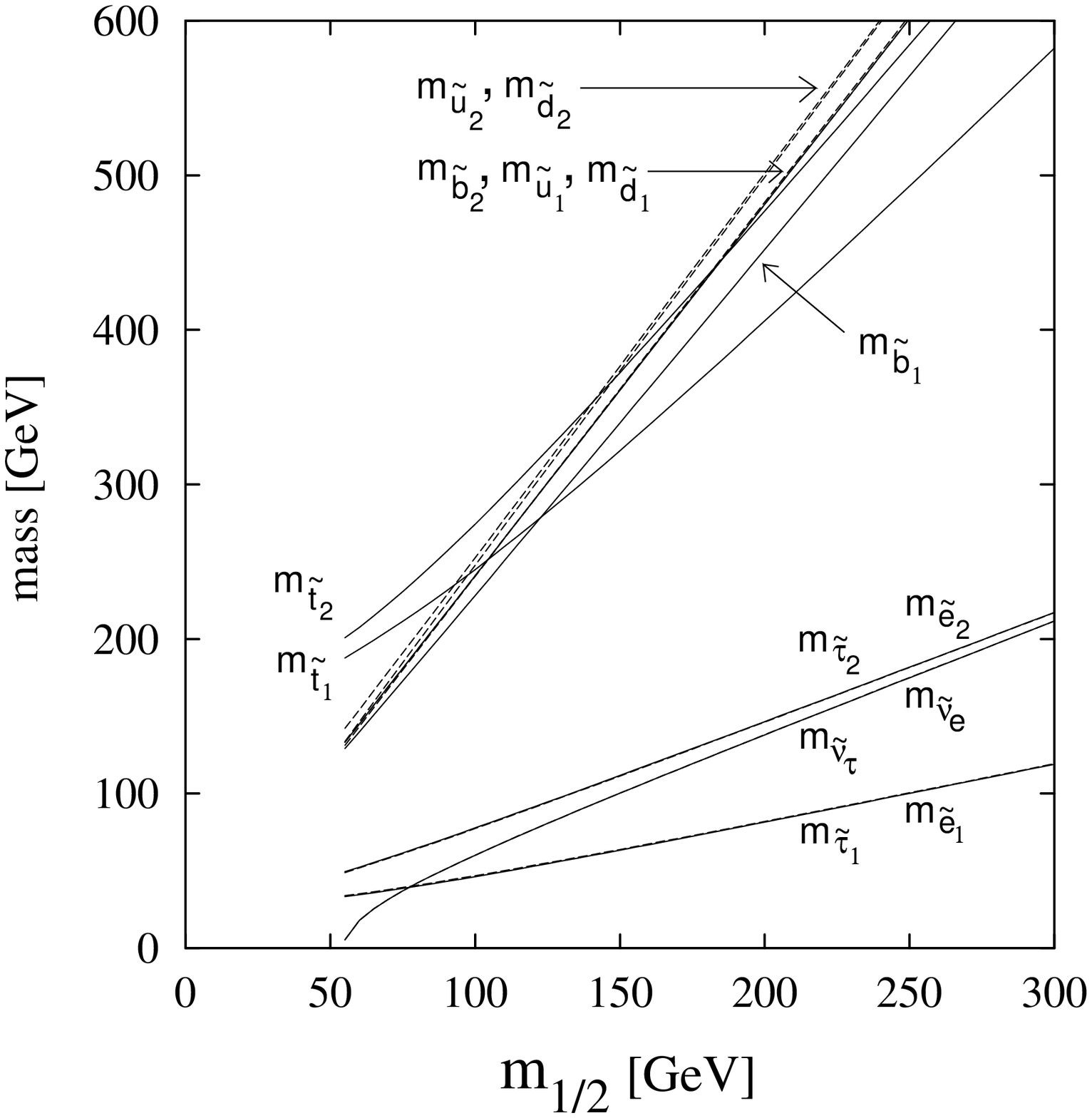}

\parbox{6in}{\small Fig.~11. Supersymmetric scalar
masses showing asymptotic behavior (from Ref.~\cite{madph801}).}
\end{center}

\begin{figure}[t]
\centering
\epsfxsize=4in\hspace*{0in}\epsffile{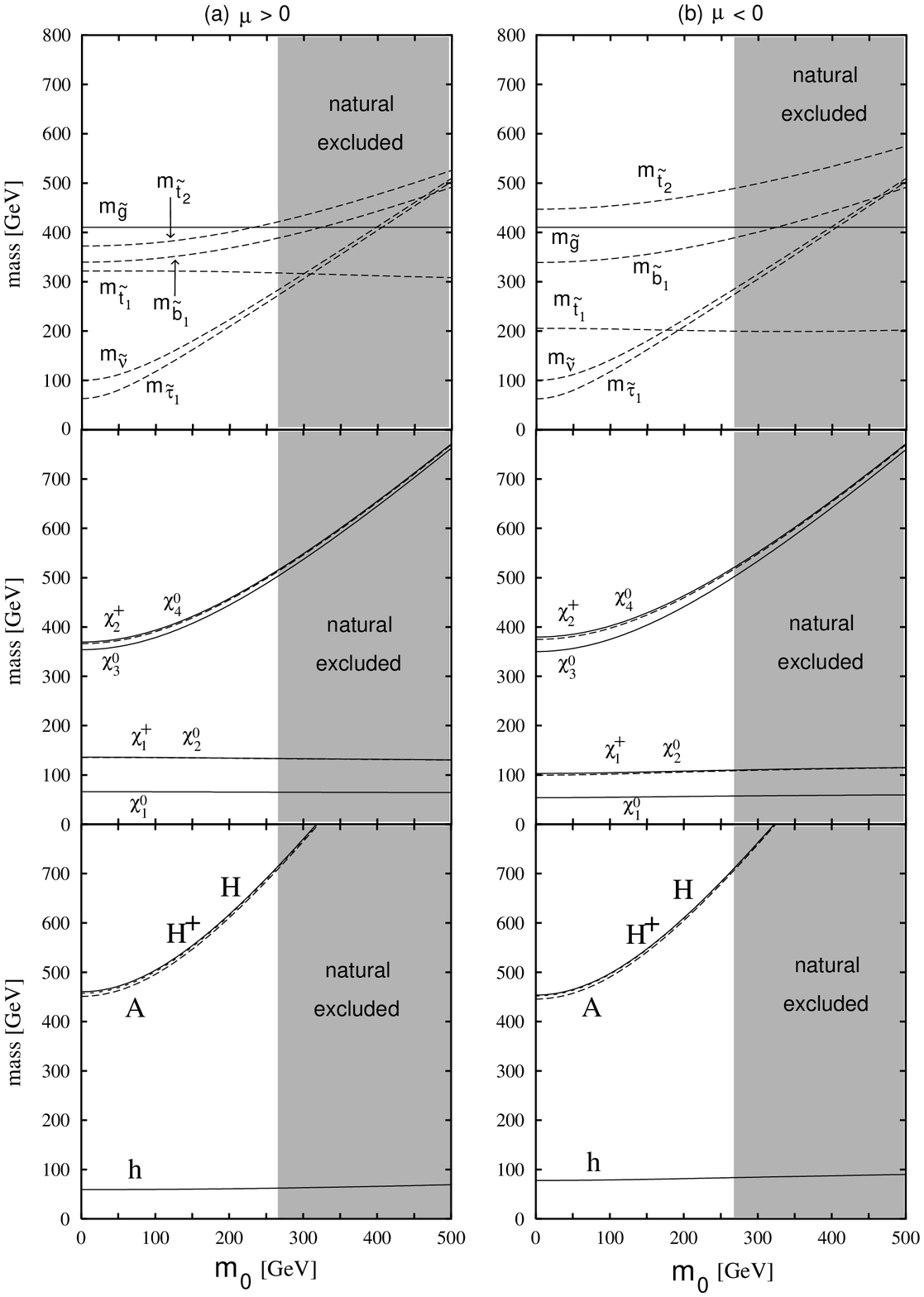}

\parbox{5.5in}{\small Fig.~12. The supersymmetric particle mass dependence on
the parameter $m_0$ for (a) $\mu > 0$ and (b) $\mu < 0$ with $m_{1/2}=150 GeV$
and $A^G=0$ (from Ref.~\cite{madph801}).}
\end{figure}

\vglue 0.6cm
{\bf\noindent 7. SUSY Signals}
\vglue 0.2cm

At hadron colliders a plethora of sparticle production processes are possible,
as illustrated in Fig.~13. The
SUSY particles must be created in pairs (for theories with an R-symmetry).
Gluinos should be copiously produced at a future hadron collider. As the
mass of the gluino increases, new decay channels open up. Figure 14 shows
typical branching fraction for gluinos assuming that
$m_{\tilde{g}}<m_{\tilde{q}}$\cite{gluinos} (in this figure it is not assumed
that $\mu $ is
large as required by radiative breaking of the electroweak symmetry).
The predicted cascade gluino decays provide multiple signals for experimental
searches. Gluino decays in the Yukawa unified supergravity scenario with
radiative electroweak symmetry breaking have been investigated in
Ref.~\cite{bdknt}.
A phenomenological discussion of the Yukawa unified no-scale model can be found
in Ref.~\cite{gunionpois}.

\begin{figure}
\centering
\epsfxsize=3.25in\hspace*{0in}\epsffile{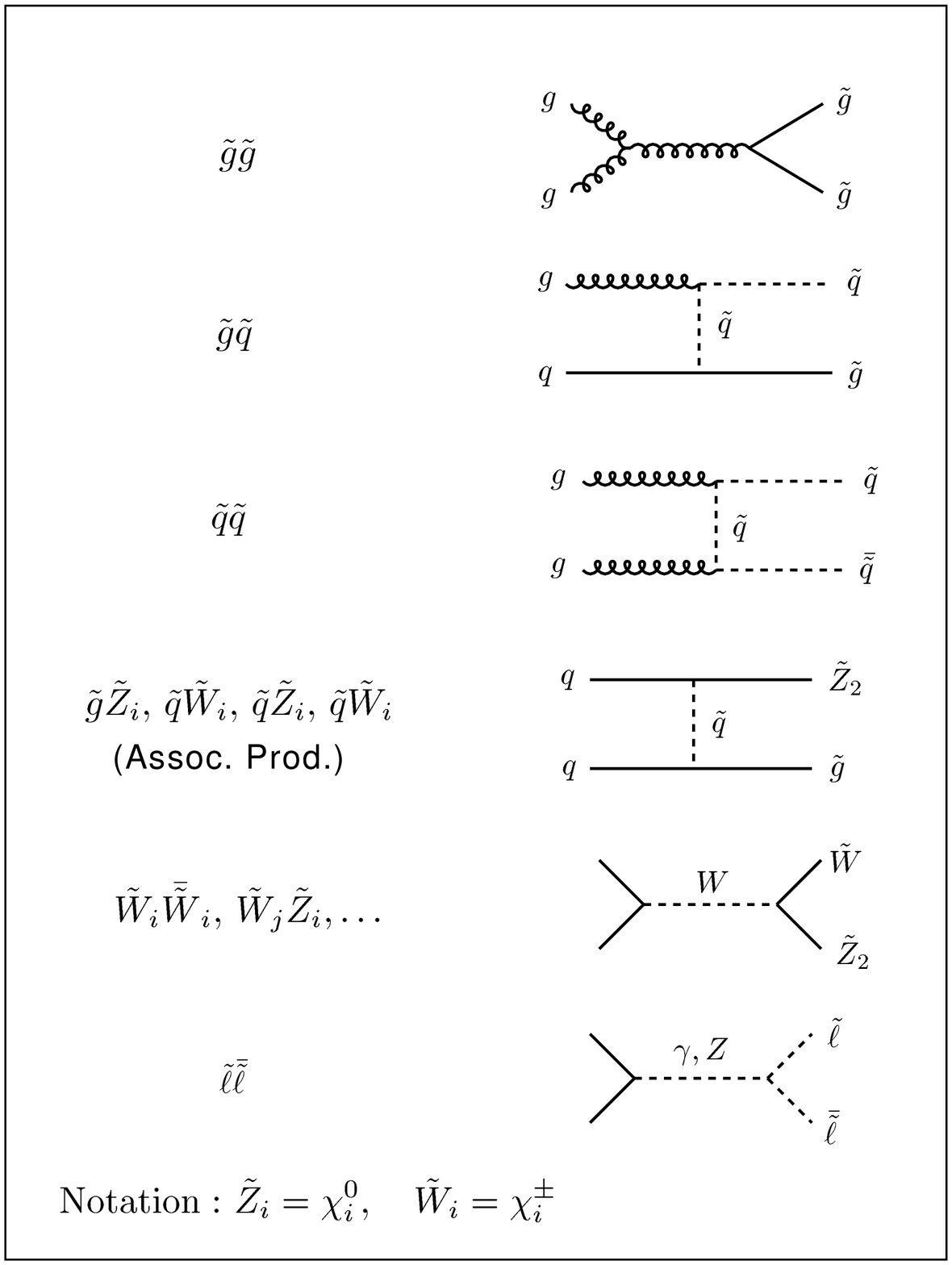}

\smallskip
{\small Fig.~13. SUSY production processes at hadron colliders.}
\end{figure}

\medskip

\begin{figure}
\centering
\epsfxsize=4in\hspace*{0in}\epsffile{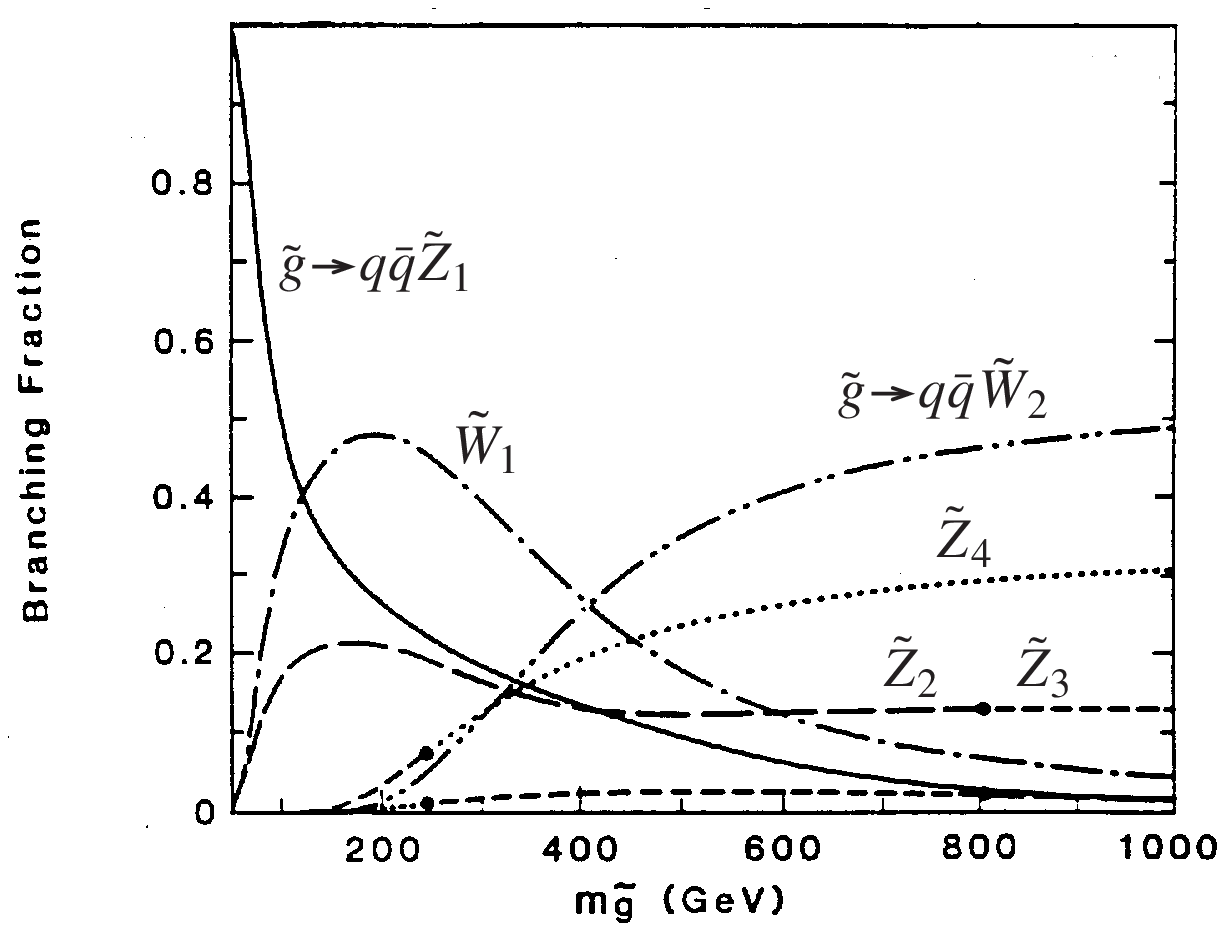}\hspace*{.25in}

\parbox{5.5in}{\small Fig.~14. Gluino branching fractions assuming that
$m_{\tilde{g}}<m_{\tilde{q}}$ (from Ref.~\cite{gluinos}).}
\end{figure}

\vglue 0.6cm
{\bf\noindent 8. Conclusions}
\vglue 0.2cm

The continued viability of supersymmetric grand unified theories with respect
to the increased precision of the low energy data calls for more
refined theoretical analyses. The following observations summarize the
principal points of this review.

\begin{itemize}

\item A low-energy supersymmetry is consistent with a desert unification
scenario in grand unified theories.

\item The observed ratio $m_b/m_{\tau}$ is consistent with SUSY GUTs.
In fact, this ratio indicates that the top quark Yukawa coupling is near
its infrared fixed point; this situation has significant implications
for SUSY Higgs searches if the top quark is lighter than about 165 GeV.
In that case the upper bound on $m_h$ is of order 100 GeV.

\item Solutions with a $\lambda _t$ fixed point,
$m_t \alt 175$ GeV and radiative breaking of the electroweak symmetry breaking
are allowed by our naturalness criterion $|\mu(M_Z)| \simeq |\mu(m_t)| < 500$
GeV for both signs of the
supersymmetric Higgs mass parameter $\mu$. These solutions are characterized by
relatively large values of $|\mu|$, which implies that the
supersymmetric particle spectrum displays a simple asymptotic behavior in the
simplest supergravity models.

\item In the early universe
the LSP annihilates sufficiently (at least in the approximation that
$s$-channel pole annihilation is neglected)
over most of the parameter space $m_0 \alt 300$ GeV, so as not to
overclose the universe.

\item The one-loop corrections to the Higgs potential somewhat
ameliorate the fine-tuning problem.

\item The tadpole method is a convenient way to calculate the
one-loop minimization conditions. We have obtained these conditions in an
analytic form including all contributions
from the gauge-Higgs sector and matter multiplets.
This method is easily extended to non-minimal Higgs sectors or to models
with additional low-lying states.

\end{itemize}

\vglue 0.6cm
{\bf\noindent Acknowledgements}
\vglue 0.2cm
This research was supported
in part by the University of Wisconsin Research Committee with funds granted by
the Wisconsin Alumni Research Foundation, in part by the U.S.~Department of
Energy under contract no.~DE-AC02-76ER00881, and in part by the Texas National
Laboratory Research Commission under grant nos.~RGFY93-221 and FCFY9302.
MSB was supported in
part by an SSC Fellowship. PO was supported in
part by an NSF Graduate Fellowship.

\vglue0.6cm
{\bf\noindent References}
\vglue 0.2cm

\end{document}